\documentclass[twocolumn,letterpaper,aps,prc,superscriptaddress,longbibliography,nofootinbib,floatfix]{revtex4-2}

\usepackage{graphicx}   
\usepackage{multirow}
\usepackage{amsmath}
\usepackage{subfigure}

\usepackage{xspace}     

\newcommand{\snn}[1]{\mbox{$\sqrt{s_{_{NN}}}=#1$\,GeV}}

\newcommand{\pp}{$p$$+$$p$\xspace}

\newcommand{\auau}{\mbox{Au$+$Au}\xspace}

\newcommand{\Ncoll}{\mbox{$N_{\rm coll}$}\xspace}

\newcommand{\gevc}{\mbox{GeV/$c$}\xspace}

\newcommand{\pt}{\mbox{$p_T$}\xspace}
\newcommand{\piz}{\mbox{$\pi^0$}\xspace}

\newcommand{\vinc}{\mbox{$v_2^{{\rm inc}}$}\xspace}
\newcommand{\vdir}{\mbox{$v_2^{{\rm dir}}$}\xspace}
\newcommand{\vdec}{\mbox{$v_2^{{\rm dec}}$}\xspace}

\newcommand{\Rg}{\mbox{$R_{\gamma}$}\xspace}

\begin{document}
 
\title{Azimuthal anisotropy of direct photons in Au$+$Au collisions at
$\sqrt{s_{_{NN}}}=200$ GeV}

\newcommand{\abilene}{Abilene Christian University, Abilene, Texas 79699, USA}
\newcommand{\augie}{Department of Physics, Augustana University, Sioux Falls, South Dakota 57197, USA}
\newcommand{\banaras}{Department of Physics, Banaras Hindu University, Varanasi 221005, India}
\newcommand{\barc}{Bhabha Atomic Research Centre, Bombay 400 085, India}
\newcommand{\baruch}{Baruch College, City University of New York, New York, New York, 10010 USA}
\newcommand{\bnlcoll}{Collider-Accelerator Department, Brookhaven National Laboratory, Upton, New York 11973-5000, USA}
\newcommand{\bnlphys}{Physics Department, Brookhaven National Laboratory, Upton, New York 11973-5000, USA}
\newcommand{\caucr}{University of California-Riverside, Riverside, California 92521, USA}
\newcommand{\charlesczech}{Charles University, Faculty of Mathematics and Physics, 180 00 Troja, Prague, Czech Republic}
\newcommand{\ciae}{Science and Technology on Nuclear Data Laboratory, China Institute of Atomic Energy, Beijing 102413, People's Republic of China}
\newcommand{\cns}{Center for Nuclear Study, Graduate School of Science, University of Tokyo, 7-3-1 Hongo, Bunkyo, Tokyo 113-0033, Japan}
\newcommand{\colorado}{University of Colorado, Boulder, Colorado 80309, USA}
\newcommand{\columbia}{Columbia University, New York, New York 10027 and Nevis Laboratories, Irvington, New York 10533, USA}
\newcommand{\czechtech}{Czech Technical University, Zikova 4, 166 36 Prague 6, Czech Republic}
\newcommand{\debrecen}{Debrecen University, H-4010 Debrecen, Egyetem t{\'e}r 1, Hungary}
\newcommand{\elte}{ELTE, E{\"o}tv{\"o}s Lor{\'a}nd University, H-1117 Budapest, P{\'a}zm{\'a}ny P.~s.~1/A, Hungary}
\newcommand{\ewha}{Ewha Womans University, Seoul 120-750, Korea}
\newcommand{\fsu}{Florida State University, Tallahassee, Florida 32306, USA}
\newcommand{\gsu}{Georgia State University, Atlanta, Georgia 30303, USA}
\newcommand{\hiroshima}{Hiroshima University, Kagamiyama, Higashi-Hiroshima 739-8526, Japan}
\newcommand{\howard}{Department of Physics and Astronomy, Howard University, Washington, DC 20059, USA}
\newcommand{\hunrenatomki}{HUN-REN ATOMKI, H-4026 Debrecen, Bem t{\'e}r 18/c, Hungary}
\newcommand{\ihepprot}{IHEP Protvino, State Research Center of Russian Federation, Institute for High Energy Physics, Protvino, 142281, Russia}
\newcommand{\illuiuc}{University of Illinois at Urbana-Champaign, Urbana, Illinois 61801, USA}
\newcommand{\inrras}{Institute for Nuclear Research of the Russian Academy of Sciences, prospekt 60-letiya Oktyabrya 7a, Moscow 117312, Russia}
\newcommand{\instpasczech}{Institute of Physics, Academy of Sciences of the Czech Republic, Na Slovance 2, 182 21 Prague 8, Czech Republic}
\newcommand{\isu}{Iowa State University, Ames, Iowa 50011, USA}
\newcommand{\jaea}{Advanced Science Research Center, Japan Atomic Energy Agency, 2-4 Shirakata Shirane, Tokai-mura, Naka-gun, Ibaraki-ken 319-1195, Japan}
\newcommand{\jeonbuk}{Jeonbuk National University, Jeonju, 54896, Korea}
\newcommand{\jyvaskyla}{Helsinki Institute of Physics and University of Jyv{\"a}skyl{\"a}, P.O.Box 35, FI-40014 Jyv{\"a}skyl{\"a}, Finland}
\newcommand{\kek}{KEK, High Energy Accelerator Research Organization, Tsukuba, Ibaraki 305-0801, Japan}
\newcommand{\korea}{Korea University, Seoul 02841, Korea}
\newcommand{\kurchatov}{National Research Center ``Kurchatov Institute", Moscow, 123098 Russia}
\newcommand{\kyoto}{Kyoto University, Kyoto 606-8502, Japan}
\newcommand{\lawllnl}{Lawrence Livermore National Laboratory, Livermore, California 94550, USA}
\newcommand{\losalamos}{Los Alamos National Laboratory, Los Alamos, New Mexico 87545, USA}
\newcommand{\lund}{Department of Physics, Lund University, Box 118, SE-221 00 Lund, Sweden}
\newcommand{\lyon}{IPNL, CNRS/IN2P3, Univ Lyon, Universit{\'e} Lyon 1, F-69622, Villeurbanne, France}
\newcommand{\maryland}{University of Maryland, College Park, Maryland 20742, USA}
\newcommand{\mass}{Department of Physics, University of Massachusetts, Amherst, Massachusetts 01003-9337, USA}
\newcommand{\mate}{MATE, Institute of Technology, Laboratory of Femtoscopy, K\'aroly R\'obert Campus, H-3200 Gy\"ongy\"os, M\'atrai \'ut 36, Hungary}
\newcommand{\michigan}{Department of Physics, University of Michigan, Ann Arbor, Michigan 48109-1040, USA}
\newcommand{\miss}{Mississippi State University, Mississippi State, Mississippi 39762, USA}
\newcommand{\muhlenberg}{Muhlenberg College, Allentown, Pennsylvania 18104-5586, USA}
\newcommand{\nara}{Nara Women's University, Kita-uoya Nishi-machi Nara 630-8506, Japan}
\newcommand{\natmephi}{National Research Nuclear University, MEPhI, Moscow Engineering Physics Institute, Moscow, 115409, Russia}
\newcommand{\newmex}{University of New Mexico, Albuquerque, New Mexico 87131, USA}
\newcommand{\nmsu}{New Mexico State University, Las Cruces, New Mexico 88003, USA}
\newcommand{\northcg}{Physics and Astronomy Department, University of North Carolina at Greensboro, Greensboro, North Carolina 27412, USA}
\newcommand{\ohio}{Department of Physics and Astronomy, Ohio University, Athens, Ohio 45701, USA}
\newcommand{\ornl}{Oak Ridge National Laboratory, Oak Ridge, Tennessee 37831, USA}
\newcommand{\orsay}{IPN-Orsay, Univ.~Paris-Sud, CNRS/IN2P3, Universit\'e Paris-Saclay, BP1, F-91406, Orsay, France}
\newcommand{\peking}{Peking University, Beijing 100871, People's Republic of China}
\newcommand{\pnpi}{PNPI, Petersburg Nuclear Physics Institute, Gatchina, Leningrad region, 188300, Russia}
\newcommand{\riken}{RIKEN Nishina Center for Accelerator-Based Science, Wako, Saitama 351-0198, Japan}
\newcommand{\rikjrbrc}{RIKEN BNL Research Center, Brookhaven National Laboratory, Upton, New York 11973-5000, USA}
\newcommand{\rikkyo}{Physics Department, Rikkyo University, 3-34-1 Nishi-Ikebukuro, Toshima, Tokyo 171-8501, Japan}
\newcommand{\saispbstu}{Saint Petersburg State Polytechnic University, St.~Petersburg, 195251 Russia}
\newcommand{\seoulnat}{Department of Physics and Astronomy, Seoul National University, Seoul 151-742, Korea}
\newcommand{\stonybrkc}{Chemistry Department, Stony Brook University, SUNY, Stony Brook, New York 11794-3400, USA}
\newcommand{\stonycrkp}{Department of Physics and Astronomy, Stony Brook University, SUNY, Stony Brook, New York 11794-3800, USA}
\newcommand{\tenn}{University of Tennessee, Knoxville, Tennessee 37996, USA}
\newcommand{\titech}{Department of Physics, Tokyo Institute of Technology, Oh-okayama, Meguro, Tokyo 152-8551, Japan}
\newcommand{\tsukuba}{Tomonaga Center for the History of the Universe, University of Tsukuba, Tsukuba, Ibaraki 305, Japan}
\newcommand{\usmma}{United States Merchant Marine Academy, Kings Point, New York 11024, USA}
\newcommand{\vandy}{Vanderbilt University, Nashville, Tennessee 37235, USA}
\newcommand{\weizmann}{Weizmann Institute, Rehovot 76100, Israel}
\newcommand{\wigner}{Institute for Particle and Nuclear Physics, HUN-REN Wigner Research Centre for Physics, (HUN-REN Wigner RCP, RMI), H-1525 Budapest 114, POBox 49, Budapest, Hungary}
\newcommand{\yonsei}{Yonsei University, IPAP, Seoul 120-749, Korea}
\newcommand{\zagreb}{Department of Physics, Faculty of Science, University of Zagreb, Bijeni\v{c}ka c.~32 HR-10002 Zagreb, Croatia}
\newcommand{\zambia}{Department of Physics, School of Natural Sciences, University of Zambia, Great East Road Campus, Box 32379, Lusaka, Zambia}
\affiliation{\abilene}
\affiliation{\augie}
\affiliation{\banaras}
\affiliation{\barc}
\affiliation{\baruch}
\affiliation{\bnlcoll}
\affiliation{\bnlphys}
\affiliation{\caucr}
\affiliation{\charlesczech}
\affiliation{\ciae}
\affiliation{\cns}
\affiliation{\colorado}
\affiliation{\columbia}
\affiliation{\czechtech}
\affiliation{\debrecen}
\affiliation{\elte}
\affiliation{\ewha}
\affiliation{\fsu}
\affiliation{\gsu}
\affiliation{\hiroshima}
\affiliation{\howard}
\affiliation{\hunrenatomki}
\affiliation{\ihepprot}
\affiliation{\illuiuc}
\affiliation{\inrras}
\affiliation{\instpasczech}
\affiliation{\isu}
\affiliation{\jaea}
\affiliation{\jeonbuk}
\affiliation{\jyvaskyla}
\affiliation{\kek}
\affiliation{\korea}
\affiliation{\kurchatov}
\affiliation{\kyoto}
\affiliation{\lawllnl}
\affiliation{\losalamos}
\affiliation{\lund}
\affiliation{\lyon}
\affiliation{\maryland}
\affiliation{\mass}
\affiliation{\mate}
\affiliation{\michigan}
\affiliation{\miss}
\affiliation{\muhlenberg}
\affiliation{\nara}
\affiliation{\natmephi}
\affiliation{\newmex}
\affiliation{\nmsu}
\affiliation{\northcg}
\affiliation{\ohio}
\affiliation{\ornl}
\affiliation{\orsay}
\affiliation{\peking}
\affiliation{\pnpi}
\affiliation{\riken}
\affiliation{\rikjrbrc}
\affiliation{\rikkyo}
\affiliation{\saispbstu}
\affiliation{\seoulnat}
\affiliation{\stonybrkc}
\affiliation{\stonycrkp}
\affiliation{\tenn}
\affiliation{\titech}
\affiliation{\tsukuba}
\affiliation{\usmma}
\affiliation{\vandy}
\affiliation{\weizmann}
\affiliation{\wigner}
\affiliation{\yonsei}
\affiliation{\zagreb}
\affiliation{\zambia}
\author{N.J.~Abdulameer} \affiliation{\debrecen} \affiliation{\hunrenatomki}
\author{U.~Acharya} \affiliation{\gsu}
\author{A.~Adare} \affiliation{\colorado} 
\author{C.~Aidala} \affiliation{\michigan} 
\author{N.N.~Ajitanand} \altaffiliation{Deceased} \affiliation{\stonybrkc} 
\author{Y.~Akiba} \email[PHENIX Spokesperson: ]{akiba@rcf.rhic.bnl.gov} \affiliation{\riken} \affiliation{\rikjrbrc}
\author{M.~Alfred} \affiliation{\howard} 
\author{S.~Antsupov} \affiliation{\saispbstu}
\author{N.~Apadula} \affiliation{\isu} \affiliation{\stonycrkp} 
\author{H.~Asano} \affiliation{\kyoto} \affiliation{\riken} 
% ...
\collaboration{PHENIX Collaboration}  \noaffiliation

\date{\today}


\begin{abstract}


The PHENIX experiment at the Relativistic Heavy Ion Collider measured 
the second Fourier component $v_2$ of the direct-photon azimuthal 
anisotropy at midrapidity in Au$+$Au collisions at 
$\sqrt{s_{_{NN}}}=200$ GeV.  The results are presented in 10\% wide bins 
of collision centrality and cover the transverse-momentum range of 
$1<p_T<20$ GeV/$c$, and are in quantitative agreement with findings 
published earlier, but provide better granularity and higher $p_T$ 
reach. Above a $p_T$ of 8--10 GeV/$c$, where hard scattering dominates 
the direct-photon production, $v_2$ is consistent with zero.  Below that 
in each centrality bin $v_2$ as a function of $p_T$ is comparable to the 
$\pi^0$ anisotropy albeit with a tendency of being somewhat smaller.  
The results are compared to recent theory calculations that include, in 
addition to thermal radiation from the quark-gluon plasma and hadron 
gas, sources of photons from pre-equilibrium, strong magnetic fields, or 
radiative hadronization. While the newer theoretical calculations 
describe the data better than previous models, none of them alone can 
fully explain the results, particularly in the region of $p_T=4$--8 
GeV/$c$.

\end{abstract}

\maketitle

\section{\label{sec:intro}Introduction}   

The quark-gluon plasma (QGP) is a deconfined state of quarks and gluons 
that arises from the principles of asymptotic freedom in quantum 
chromodynamics, the theory of strong interactions. Since the 
establishment of the Relativistic Heavy Ion Collider (RHIC) in the early 
2000s, the formation of QGP has been well-documented in relativistic 
heavy-ion 
collisions~\cite{BRAHMS:2004adc,PHENIX:2004vcz,PHOBOS:2004zne,STAR:2005gfr}. 
Photons serve as a unique probe for studying the space-time evolution of 
these collisions~\cite{shuryak} as their mean free path is larger than 
the system size and, hence, carry unaltered information about the 
dynamics of the system as it explosively expands and cools. The photons 
of interest, those that do not originate from hadron decays, are referred 
to as direct photons. Sources of direct photons, ordered by emission 
time, include (a) prompt photons produced in early hard-scattering 
processes~\cite{Aurenche:1983ws,Owens:1986mp,Paquet:2015lta}, (b) a number 
of possible mechanisms emitting photons before QGP is formed, such as
pre-equilibrium phase~\cite{Gale:2021emg}, the hot-glue 
scenario~\cite{Shuryak:1992bt}, the Glasma 
state~\cite{McLerran:2014hza}, the strong magnetic field created in 
off-central collisions~\cite{Tuchin:2010gx} etc, and (c) radiation from the 
fireball of the QGP phase through hadronization until kinetic 
freezeout of 
hadrons~\cite{Linnyk:2015rco,Paquet:2015lta,Shen:2013vja,vanHees:2011vb,Dusling:2009ej,Turbide:2003si,Dion:2011pp,Gale:2021emg}; 
and final-state effects, such as jet-medium 
interactions~\cite{Renk:2013kya,Fries:2002kt}.

Measurements of direct photons at 
RHIC~\cite{PHENIX:2008uif,PHENIX:2011oxq,PHENIX:2012jbv,PHENIX:2014nkk,PHENIX:2015igl,STAR:2016use,PHENIX:2018for,PHENIX:2022qfp,PHENIX:2022rsx} 
and the Large Hadron Collider~\cite{ALICE:2015xmh,ALICE:2018dti} 
constrain initial conditions, sources of photon production, emission 
rates, and the space-time evolution of relativistic heavy-ion 
collisions.  In general, theoretical models are qualitatively consistent 
with the observed large yields, azimuthal anisotropy, and centrality 
dependence. However, they do not describe the details of the data 
quantitatively (for a review see~\cite{David:2019wpt}).

This study focuses on the azimuthal anisotropy of direct-photon 
emission, which is highly sensitive to the initial state of the 
fireball. Even slight modifications can lead to significant variations 
in the final-state results~\cite{Chatterjee:2024blo}. Anisotropies arise 
in off-center heavy-ion collisions, which create an elliptical 
interaction region, generating pressure gradients along the surface. 
These pressure gradients drive an anisotropic expansion of the 
fireball~\cite{Heinz:2013th}. Photons emitted from the fireball will 
experience an anisotropic blue shift~\cite{vanHees:2011vb,Linnyk:2015rco,Paquet:2015lta,Shen:2013vja,Chatterjee:2008tp,Campbell:2015jga,Fujii:2022hxa}, depending on the time 
of emission. Prompt and early photons will be largely unaffected, while 
photons from the expanding hadron gas (HG) will experience the largest 
blue shifts and thus the largest anisotropies. Other sources of 
anisotropies may exist. For example, the strong, polarized 
electromagnetic fields created during heavy-ion collisions might alter 
emission rates early in the 
collisions~\cite{Basar:2012bp,Muller:2013ila,Wang:2020dsr,Sun:2024vsz}. 
These changes will depend on the direction relative to the polarization 
of the fields, thereby creating an anisotropy with respect to the 
reaction plane. Also photon emission influenced by final-state effects, 
such as jet quenching, will be altered differently depending on the path 
length through the medium; again resulting in an 
anisotropy~\cite{Renk:2013kya,Fries:2002kt}.

The quantity measured, in this context, is the Fourier transform of the 
azimuthal distribution of 
photons~\cite{Voloshin:1994mz,Poskanzer:1998yz}, which can 
be written as

\begin{equation}
\frac{dN}{dp_T d(\phi-\Psi_k)} \propto 1 + \sum_n v_{n,k} \cos(n(\phi-\Psi_k)),
\label{eqn:flow}
\end{equation}

\noindent where $\phi$ is the azimuthal angle in the global coordinate 
system, and $\Psi_k$ is the orientation of the $k$-th order event plane 
in the given event, estimated from the distribution of particles in a 
forward event-plane detector. The coefficients $v_{n,k}$ are the $n$-th 
harmonic with respect to the $k$-th reaction plane. Except for the $k=2$ 
event plane, which is to first order the reaction plane, all other 
planes result from density fluctuations in the incoming beams. 
Statistics limits this study to the $k=2$ event plane with the leading 
coefficient $v_{2,2} \equiv v_2\{\Psi_2\}$, which we will refer to as 
$v_2$, and a minor contribution from $v_{4,2} \equiv v_4\{\Psi_2\}$, 
which we will refer to as $v_4$.

In this paper, the PHENIX collaboration reports the measurement of $v_2$ 
for direct photons in Au$+$Au collisions at RHIC at 
$\sqrt{s_{_{NN}}}=200$~GeV using the high-statistics data collected in 
2014. Section~\ref{sec:exp} describes the experimental setup, followed 
by data analysis in Section~\ref{sec:ana}. Sources of systematic 
uncertainties are presented in Section~\ref{sec:sys}, and results are 
discussed in Sections~\ref{sec:result} and~\ref{sec:discussion}.  
Finally, a summary is given in Section~\ref{sec:summ}

\section{\label{sec:exp}Experimental setup and photon measurements}  

\begin{figure}[ht!]
     \includegraphics[width=1.0\linewidth]{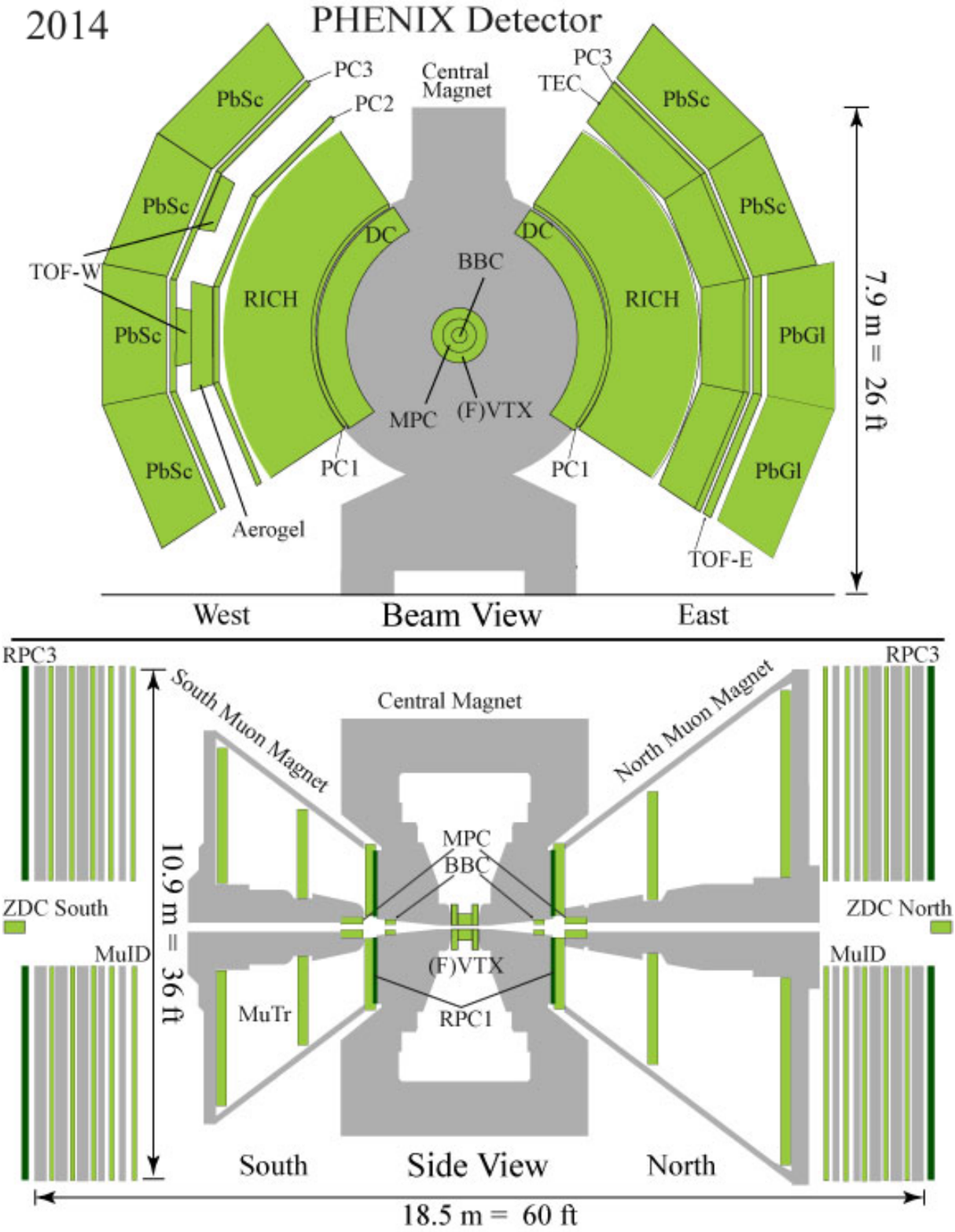}
\caption{Beam view and side view of the PHENIX detector setup for the 
year 2014.}
   \label{fig:phenix}
\end{figure}

In 2014, PHENIX recorded $1.9 \times 10^{10}$ Au$+$Au collisions or 
events at $\sqrt{s_{_{NN}}}=200$ GeV with the detector setup shown in 
Fig.~\ref{fig:phenix}. Of these $1.25 \times 10^{10}$ events passed all 
quality and selection cuts in the present analysis. Collisions were 
identified by a minimum-bias trigger that requires coincident 
signals in two beam-beam counters (BBC)~\cite{PHENIX:BBC} that are 
located on either side of the interaction point along the beam axis at 
$z={\pm}1.44$~m. Each BBC is segmented into 64 \v{C}erenkov counters 
that measure the number of produced charged particles and their arrival 
time in $3.1<|\eta|<3.9$. The timing information provides the 
trigger and the $z-$position of the collision vertex along the beam 
direction with a resolution of less than a centimeter. Only events that 
occurred within $\pm10$~cm of the nominal interaction point are 
analyzed. The particle multiplicity information is used to divide events 
into 10\% centrality classes.

The direct photon measurement is performed using the electromagnetic 
calorimeters (EMCal)~\cite{PHENIX:EMCal} in the central-arm 
spectrometers. Each arm has 4 sectors covering a total of $2 \times 
90^{\circ}$ in azimuth and $|\eta| < 0.35$. Two types of calorimeters 
are used: 6 sectors are lead-scintillator (PbSc) sampling EMCal and 2 
are lead-glass (PbGl) EMCal, which provide energy resolutions of 
$\sigma_E/E=8.1\%/\sqrt{E~{\rm [GeV]}}{\oplus}5.0\%$, and 
$\sigma_E/E=8.7\%/\sqrt{E~{\rm [GeV]}}{\oplus}5.8\%$, respectively.  
Photon identification is accomplished through cuts on shower shape, 
comparing the reconstructed showers to template electromagnetic showers. 
A match of $\chi^2<3$ is required. The selected photon sample 
covers a range of $0.5<\pt<20$~\gevc.

The event plane, $\Psi_2$, is measured with the two forward-vertex 
(FVTX) detectors~\cite{PHENIX:FVTX} of the PHENIX silicon-vertex 
tracker. The FVTX extends the vertex tracking to the range of 
$1.2<|\eta|<2.2$ (for tracks) and $1.0<|\eta|<3.0$ (for clusters) and 
2$\pi$ in azimuth. Each side has 4 stations in the beam direction with 48 
individual silicon sensors. Each sensor contains two columns of 
ministrips with 75~$\mu{\rm m}$ pitch in the radial direction and 
lengths in the azimuthal direction varying from 3.4~mm at the inner 
radius to 11.5~mm at the outer radius. The FVTX has approximately 0.54 
million strips on each side. $\Psi_2$ can be measured with a resolution 
of up to $75\%$ for semicentral events.

\section{\label{sec:ana}Data Analysis}  

The azimuthal anisotropy of direct photons \vdir with respect to the 
event plane $\Psi_2$ is calculated from the measured anisotropy of all 
photon candidates \vinc and that of modeled photons from hadron decays 
\vdec.  Using the fraction \Rg of the inclusive photon yield divided by 
the yield of decay photons, the direct-photon anisotropy can be 
calculated as

\begin{equation}
v_2^{{\rm dir}} = \frac{R_\gamma v_2^{{\rm inc}} - v_2^{{\rm dec}}}{R_\gamma - 1}.
\label{eqn:vdir}
\end{equation}

\noindent where all quantities are functions of photon \pt. 
Each term in Eq.~\ref{eqn:vdir} and $\Psi_2$ are determined separately.

\subsection{The event plane $\Psi_2$}

The reaction plane of a collision is the plane defined by the 
intersection of the beam direction with the impact parameter of the two 
colliding nuclei. However, this is experimentally inaccessible. It is 
instead estimated using the event plane 
$\Psi_2$~\cite{Poskanzer:1998yz,Ollitrault:1992bk} measured from the 
distribution of particles detected in the FVTX.

The orientation of the event plane $\Psi_2$ in global coordinates is 
calculated using the azimuthal distribution of the charged particle 
tracks reconstructed in the FVTX for a given event. The components of 
the event-plane vector $Q$ are defined as $Q_x$ and $Q_y$:

\begin{equation}
Q_x = \sum_i \cos(2\phi_i), \quad Q_y = \sum_i \sin(2\phi_i),    
\end{equation}

\noindent where ($\phi_i$) is the azimuthal angle of the ($i$-th) 
particle. The event-plane $\Psi_2$ is then calculated as:

\begin{equation}
\Psi_2 = \frac{1}{2} \arctan\left(\frac{Q_y}{Q_x}\right). 
\end{equation}

\begin{figure}[ht!]
     \includegraphics[width=1.0\linewidth]{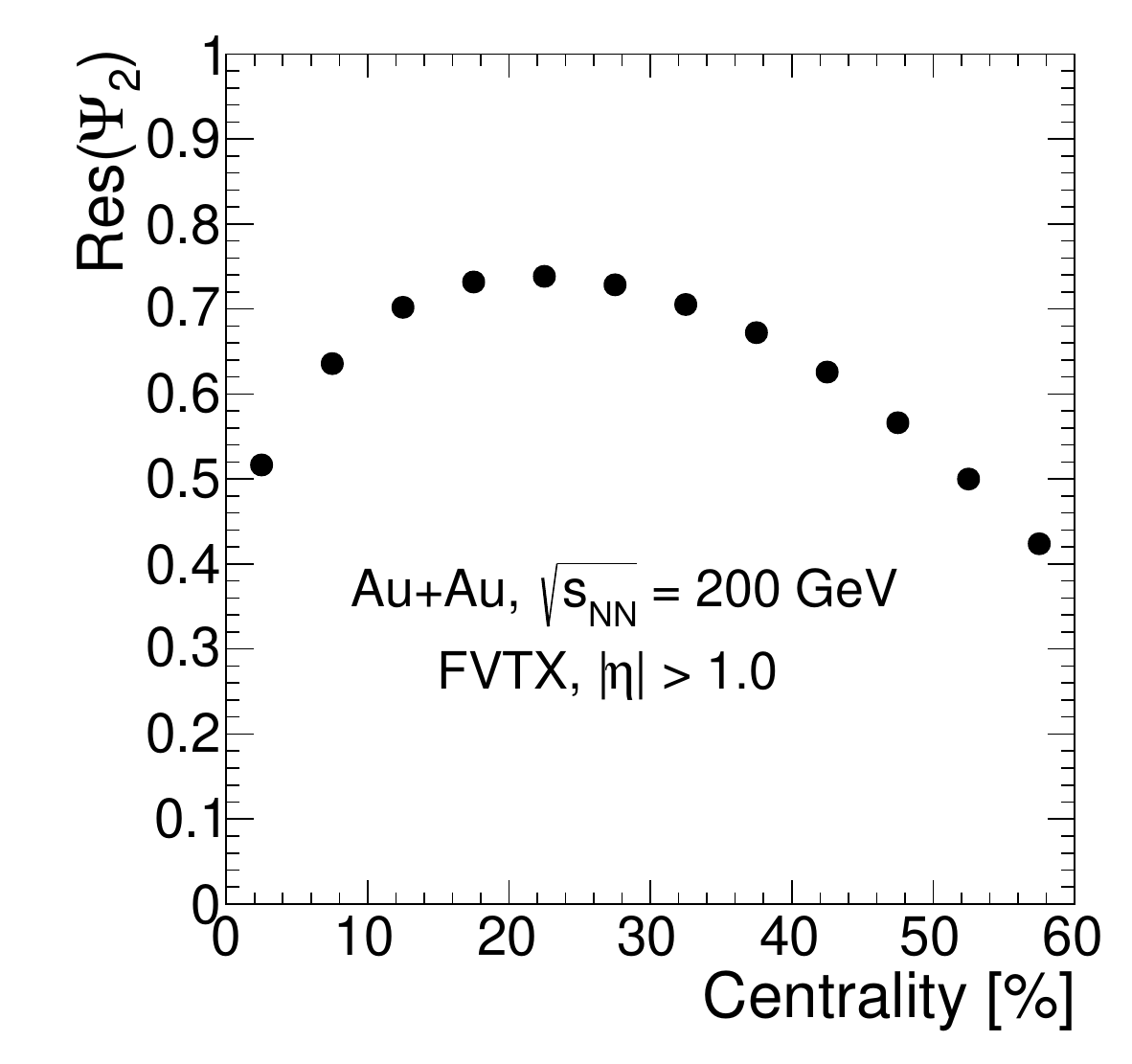}
\caption{The event-plane resolution for different centralities using the 
three subevent method.}
   \label{fig:ep_res}
\end{figure}


The event-plane resolution $\rm Res(\Psi_2)$ is calculated using the 
combinations of three subevents. In this approach, the event-plane 
angles from three independent detector regions are compared. The 
resolution factor is given by

\begin{equation}
{\rm Res(\Psi_2)}=\sqrt{
\frac{\langle \cos(2[\Psi_2^{A}-\Psi_2^{B}]) \rangle \langle \cos(2[\Psi_2^{A}-\Psi_2^{C}]) \rangle}{\langle\cos(2[\Psi_2^{B}-\Psi_2^{C}])\rangle}},
\end{equation}

\noindent where $\Psi_2^{A}, \Psi_2^{B}$, and $\Psi_2^{C}$ are the event 
planes determined from (A) tracks in the FVTX, (B) hits in the BBC 
detector, and (C) tracks reconstructed in the central-arm detectors. The 
event-plane resolution is shown in Fig.~\ref{fig:ep_res} as a function 
of centrality. The event-plane resolution depends on the number of 
charged particles produced in a collision and the size of the anisotropy 
of the charged-particle production for a given centrality class. This 
leads to a broad maximum near semicentral collisions, where $\rm 
Res(\Psi_2)$ reaches $\approx$75\%. Towards more central collisions it 
reduces to $\approx$50\%; and as collisions become more peripheral the 
resolution drops to $\approx$40\%.

\subsection{Inclusive photon anisotropy \vinc}

Showers in the EMCal are identified as photon candidates by comparing 
the shower shape to templates of electromagnetic showers using a 
$\chi^2<3$ match. For a given centrality and \pt-range, the anisotropy 
of all photon candidates is measured directly by fitting the 
distribution of azimuthal angles, $\phi$, of the photons relative to the 
event plane $\Psi_2$. The functional form of the fit is given as

\begin{equation}
\frac{dN}{d\Delta\phi} =  \left<\frac{dN}{d\phi}\right>\left(1 + 2v_2\cos(2\Delta\phi) + 2v_4\cos(4\Delta\phi)\right),
\label{eqn:v2fit}
\end{equation}

\begin{figure}[ht!]
    \centering
    \includegraphics[width=1.0\linewidth]{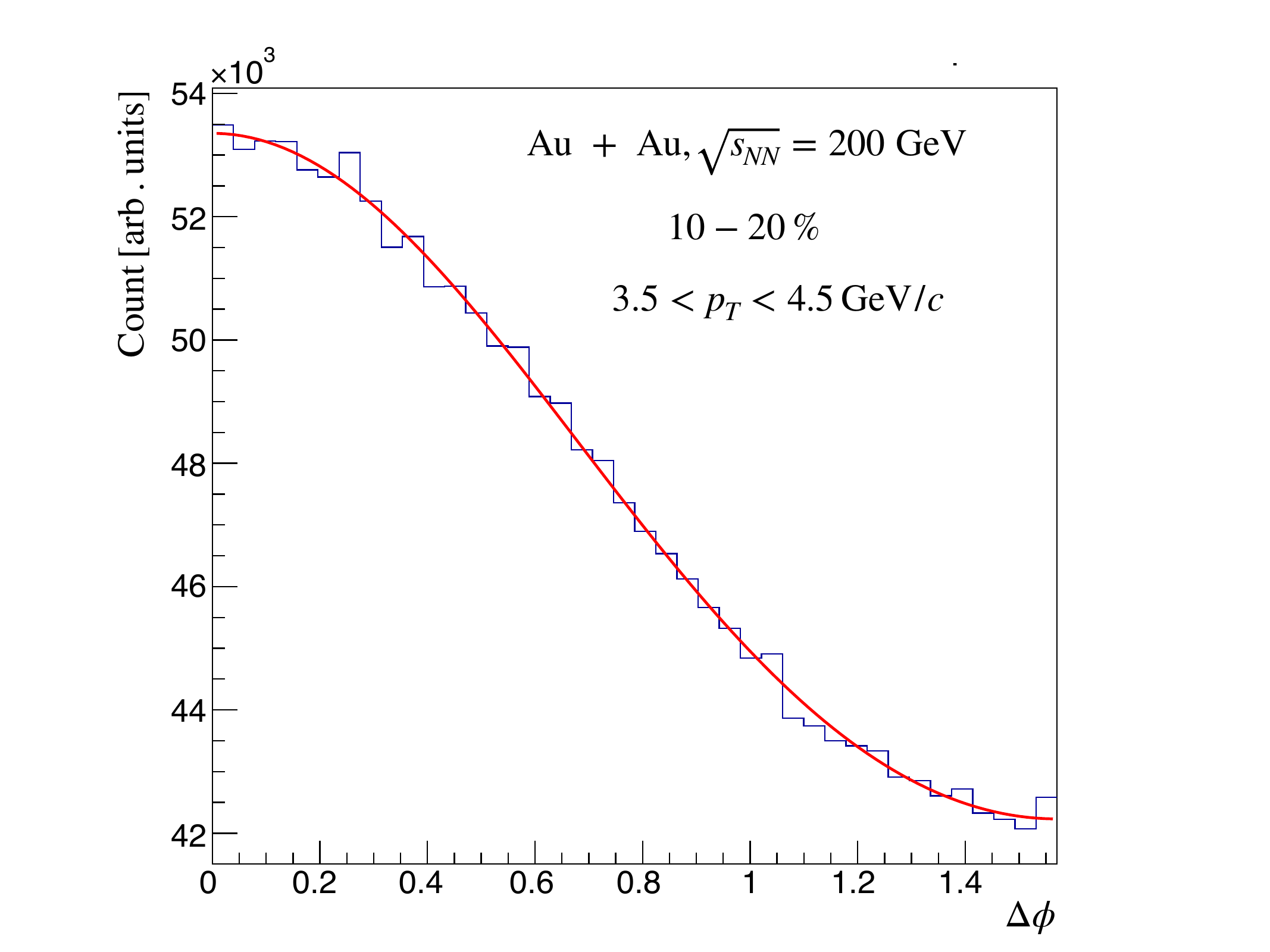}
    \caption{Sample plot of $dN/d\Delta\phi$ vs $\Delta \phi$. The red 
line is the cosine fit used to determine $v_2$.}
    \label{fig:inc_fit}
\end{figure}

\noindent where $\Delta\phi=\phi-\Psi_2$. An example is shown in 
Fig.~\ref{fig:inc_fit}. The extracted parameter $v_2$ underestimates 
the actual \vinc because of the finite event-plane resolution. To obtain 
\vinc, $v_2$ needs to be corrected with
\begin{equation}
\vinc=\frac{v_2}{\rm Res (\Psi_2)}.
\end{equation}

\subsection{Decay photon anisotropy \vdec}

The modeling of \vdec is based on a fast simulation of \piz, $\eta$, 
$\omega$, and $\eta'$ using the PHENIX decay generator, which  
simulates mesons according to given input \pt spectra; decays them based 
on the known decay kinematics and branching ratios; and aggregates the 
decay photons in the PHENIX-detector acceptance.

The mesons are generated with $0.1 < \pt < 25$ \gevc, $|y| < 0.5$ and 
$2\pi$ in azimuth. The \pt distribution for \piz is based on 
experimental data ~\cite{PHENIX:2003iij,PHENIX:2003qdj,PHENIX:2008saf} 
fitted by a modified Hagedorn function:
\begin{equation}
     \frac{1}{2\pi p_T} \frac{dN}{dp_T} = A \left( e^{-a m_T - b m_T^2} + \frac{m_T}{p_0} \right)^{-n}, 
    \label{eqn:hagedorn}
\end{equation}
\begin{equation}
     m_T = \sqrt{p_T^2 + m^2 - m_{\pi^0}^2},
     \label{eqn:mtscaling}
\end{equation}

\noindent where the parameters $A$, $a$, $b$, $p_0$, $n$ for \piz are 
given in Table~\ref{tab:hagedorn}.

\begin{table}[htb]  
\caption{Parameters for \piz for the modified Hagedorn function 
Eq.~\ref{eqn:hagedorn} to PHENIX data~\cite{PHENIX:2003iij, 
PHENIX:2003qdj, PHENIX:2008saf} from \auau collisions at \snn{200}. 
} 
{\label{tab:hagedorn}}
\begin{ruledtabular} \begin{tabular}{cccccc} 
 centrality   & $A$                & $a$                & $b$                & $p_0$   & $n$  \\
              & (GeV/$c$)$^{-2}$ & (GeV/$c$)$^{-1}$ & (GeV/$c$)$^{-2}$ & GeV/$c$ &    \\
\hline
    min.bias  & 504.5            & 0.5169           & 0.1626           & 0.7366  & 8.274  \\
    0\%--10\%    & 1331.0           & 0.5654           & 0.1945           & 0.7429  & 8.361  \\
    10\%--20\%   & 1001.0           & 0.5260           & 0.1628           & 0.7511  & 8.348  \\
    20\%--30\%   & 750.7            & 0.4900           & 0.1506           & 0.7478  & 8.229  \\
    30\%--40\%   & 535.3            & 0.4534           & 0.1325           & 0.7525  & 8.333  \\
    40\%--50\%   & 364.5            & 0.4333           & 0.1221           & 0.7385  & 8.261  \\
    50\%--60\%   & 231.2            & 0.4220           & 0.1027           & 0.7258  & 8.220  \\
    60\%--70\%   & 118.1            & 0.4416           & 0.0559           & 0.7230  & 8.163  \\
    70\%--80\%   & 69.2             & 0.2850           & 0.0347           & 0.7787  & 8.532  \\
    80\%--93\%   & 51.1             & 0.2470           & 0.0619           & 0.7101  & 8.453  \\
\end{tabular} \end{ruledtabular}
\end{table} 

For $\eta$, the \pt spectra are derived by scaling the \piz spectra with 
the $\eta$/\piz ratio, as detailed in~\cite{Ren:2021xbh}

\begin{equation}
E\frac{d^{3}N_{\eta}} {dp^3} = E\frac{d^{3}N_{\pi^{0}}} {dp^3} \times \eta/\pi^{0} \times R_{\rm flow},    
\end{equation}

\noindent where $R_{\rm flow}$ accounts for radial-flow effects, calculated 
as the ratio of $K^{\pm}/\pi^{\pm}$ in \auau collisions to that in \pp 
collisions. This method incorporates world data for $\eta$/\piz and 
avoids the over-estimation from $m_T-$scaling. The momentum 
distributions for $\omega$ and $\eta^{\prime}$ are obtained from 
$m_T$-scaling using Eq.~\ref{eqn:mtscaling}. The normalization of 
$\omega$ and $\eta^{\prime}$ are fixed at \pt= 5~\gevc to $0.9\pm0.06$ 
and $0.25\pm0.075$, respectively~\cite{PHENIX:2014nkk}.

The \pt distributions are modulated with $v_2$. For \piz the $v_2$ is 
based on experimental data from \auau collisions at \snn{200} for 
charged pions~\cite{PHENIX:2014uik} and neutral 
pions~\cite{PHENIX:2015idk}. An empirical functional form, given by

\begin{equation}
    \frac{v_2}{n_q} = N_1 \tan^{-1}(\alpha x) + N_2 (x^2 + \beta x) e^{-\lambda x}, 
\end{equation}
\begin{equation}
    x = \frac{1}{n_q} \left(\sqrt{p_T^2 + m^2} - m \right) = KE_T,
\end{equation}

\noindent is fitted to the data, where the fit parameters are $N_1$, $\alpha$, 
$N_2$, $\beta$, and $\lambda$; $n_q$ is the number of valence quarks in the 
particle ($n_q = 2$ for mesons); $KE_T$ is the transverse kinetic 
energy; and $m$ is the particle mass. The shape was chosen so that $v_2$ 
is zero at $\pt=0$ and a constant asymptotic value at high \pt, while 
allowing for a maximum at medium \pt. An example is shown in 
Fig.~\ref{fig:dec_fit}. Here and in all following figures statistical 
uncertainties are shown as error bars, while systematic uncertainties 
are shown as shaded bars or bands.

\begin{figure}[ht!]
    \centering
    \includegraphics[width=1.0\linewidth]{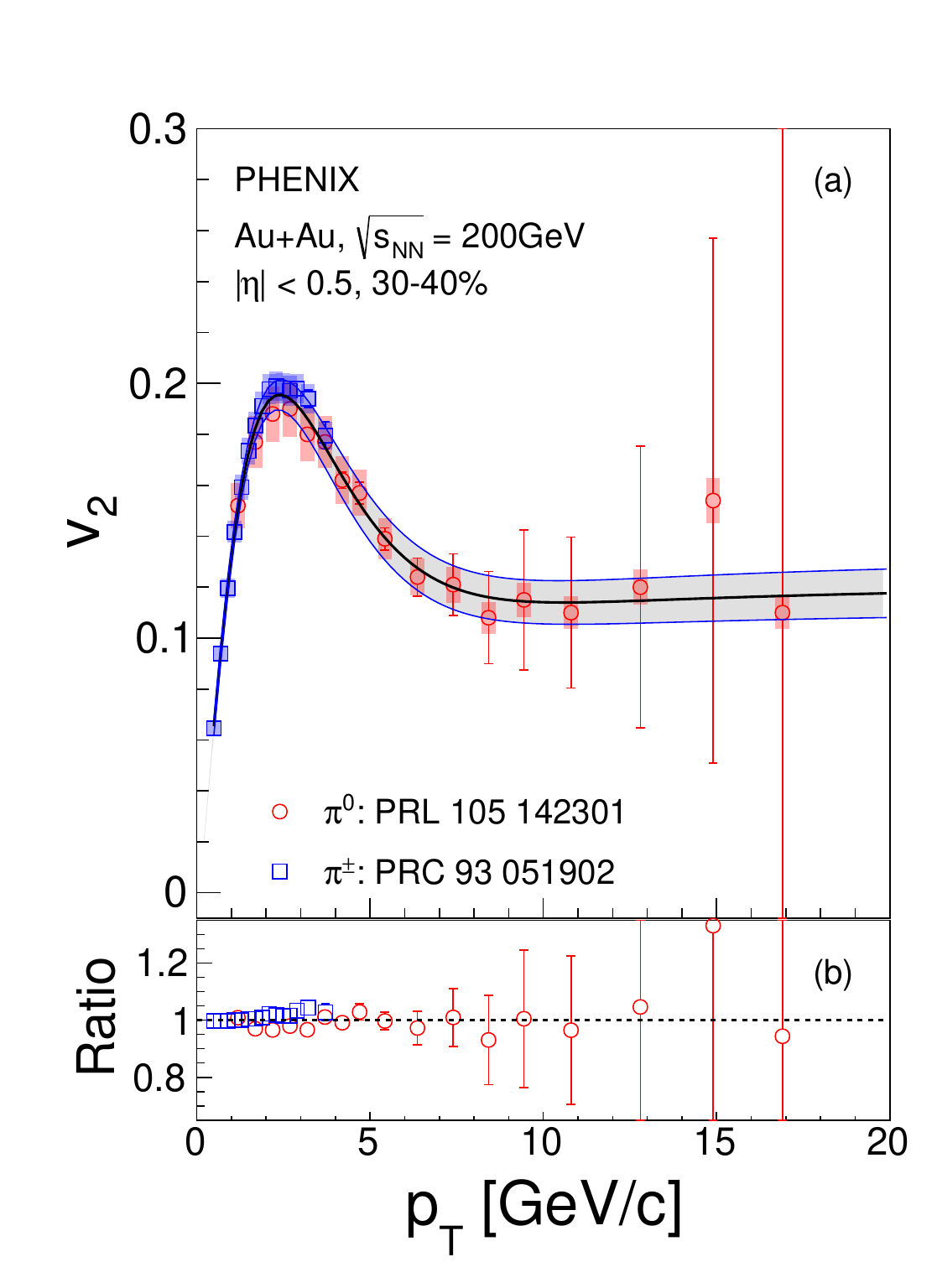}
    \caption{Combined fit for neutral and charged pion $v_2$ data sets 
as a function of \pt; statistical errors are shown as bars, systematic 
uncertainties as shaded area. The upper and lower bounds of the fit, 
shown as band, are determined by an MC sampling method.}
    \label{fig:dec_fit}
\end{figure}

The $v_2$ of heavier mesons is assumed to follow the same functional 
form as pions, scaled with $KE_T$ according to mass 
differences~\cite{PHENIX:2014uik}:

\begin{equation}
    v_2(KE_T^{\pi^0}) = v_2(KE_T^{\eta, \omega, \eta^{\prime}}).
\end{equation}

In the next step, all mesons are decayed to final states containing 
photons, weighted by their branching ratios. The photons are aggregated 
in bins of \pt and for each bin the $dN/d\phi$ distribution is fitted to 
determine \vdec.

\subsection{Determination of \Rg}

\begin{figure}
    \centering
    \includegraphics[width=1.0\linewidth]{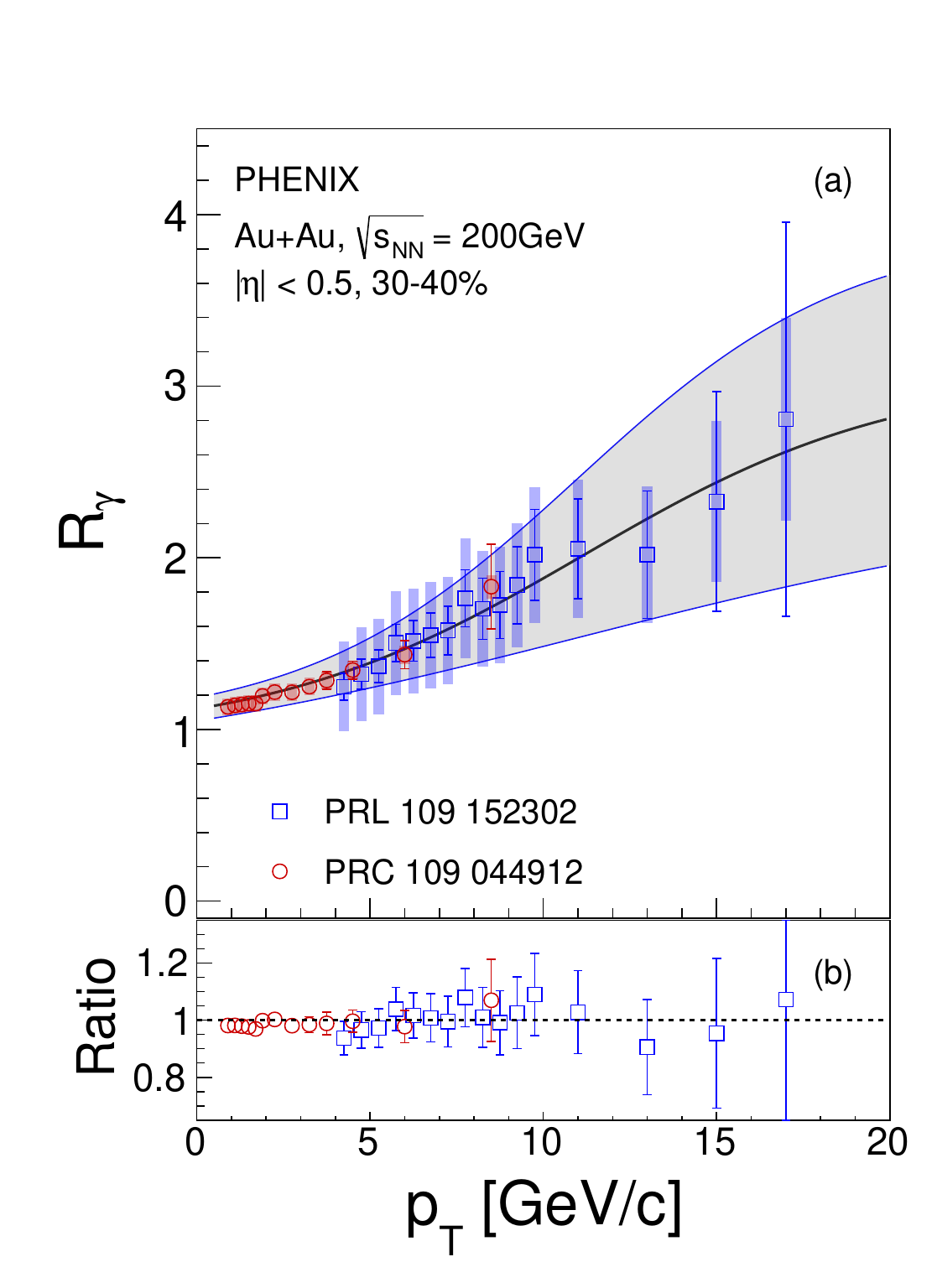}
    \caption{Sample plot of \Rg vs. \pt; statistical errors of the data 
are shown as bars, systematic uncertainties as shaded area. The fit, 
obtained using the Monte Carlo sampling procedure, is given with 
systematic uncertainties shown as band.}
    \label{fig:rgamma}
\end{figure}

Finally, the ratio \Rg, which is the inclusive photon yield divided by 
the photon yield from hadron decays, is needed to appropriately weight 
the inclusive-photon contribution in Eq.~\ref{eqn:vdir}. The ratio \Rg 
is determined from a combined fit to data previously published by PHENIX 
for \auau collisions at \snn{200}~\cite{PHENIX:2012jbv,PHENIX:2022rsx}. 
The data for 10\% centrality bins are fitted with an empirical 
functional form:

\begin{equation}
     R_\gamma (p_T) = M_0 + \frac{M_1}{1 + e^{m_0(p_T - m_1)}},
\end{equation}

\noindent where $M_0$, $M_1$, $m_0$, and $m_1$ are the fit parameters. A 
representative plot of \Rg versus \pt, along with the fitted curve and 
systematic uncertainties, is shown in Fig. \ref{fig:rgamma}. The 
functional form has no direct physical meaning, but was chosen to give a 
good representation of the data with the following features. The value 
of \Rg is approximately 1.15 at \pt$\approx$1~\gevc, for all centrality 
ranges studied here. \Rg then increases monotonically over the entire \pt 
range and saturates at very high \pt. The value at \pt~$\approx$18~\gevc 
increases with centrality, consistent with the larger suppression of 
\piz in more central collisions and the absence of a nuclear 
modification for prompt photons.

\begin{table}[htb]
\caption{Absolute systematic uncertainties on \vdir. Values shown are 
selected \pt and centralities. Positive and negative uncertainties are 
averaged here, but shown separately in 
Figs.~\ref{fig:dir_results_10}--\ref{fig:model}. In addition to the 
total uncertainty on \vdir, also given are the individual contributions 
due to uncertainties on \Rg, \vinc, and \vdec, which are calculated 
using the same sampling method but setting the other uncertainties to 
zero.
}
\label{dir_v2_sys_summary} 
\begin{ruledtabular} \begin{tabular}{cccccc}
{Uncertainty} & {centrality} & \multicolumn{4}{c}{\pt [\gevc]}  \\
on \vdir due to       &        &  2.5  & 5      & 10    & 15      \\
\hline
{total}  & 0\%--10\%       & 0.023 & 0.018  & 0.015 & 0.025   \\
         & 30\%--40\%      & 0.05  & 0.03   & 0.035 & 0.06    \\
         & 50\%--60\%      & 0.08  & 0.08   & 0.06  & 0.09    \\ \\
\Rg      & 0\%--10\%       & 0.003 &  0.003 & 0.003 & 0.005   \\
         & 30\%--40\%      & 0.01  &  0.01  & 0.015 & 0.015   \\
         & 50\%--60\%      & 0.025 &  0.025 & 0.025 & 0.025   \\ \\
\vinc    & 0\%--10\%       & 0.016 &  0.016 & 0.014 & 0.023 \\
         & 30\%--40\%      & 0.04  &  0.024 & 0.03  & 0.06  \\
         & 50\%--60\%      & 0.035 &  0.037 & 0.04  & 0.08  \\ \\
\vdec    & 0\%--10\%       & 0.018 &  0.01  & 0.005 & 0.005 \\
         & 30\%--40\%      & 0.037 &  0.019 & 0.011 & 0.009 \\
         & 50\%--60\%      & 0.09  &  0.07  & 0.05  & 0.03  \\ 
\end{tabular} \end{ruledtabular} 
\end{table}

\begin{figure}[htb]
    \centering
    \includegraphics[width=1.0\linewidth]{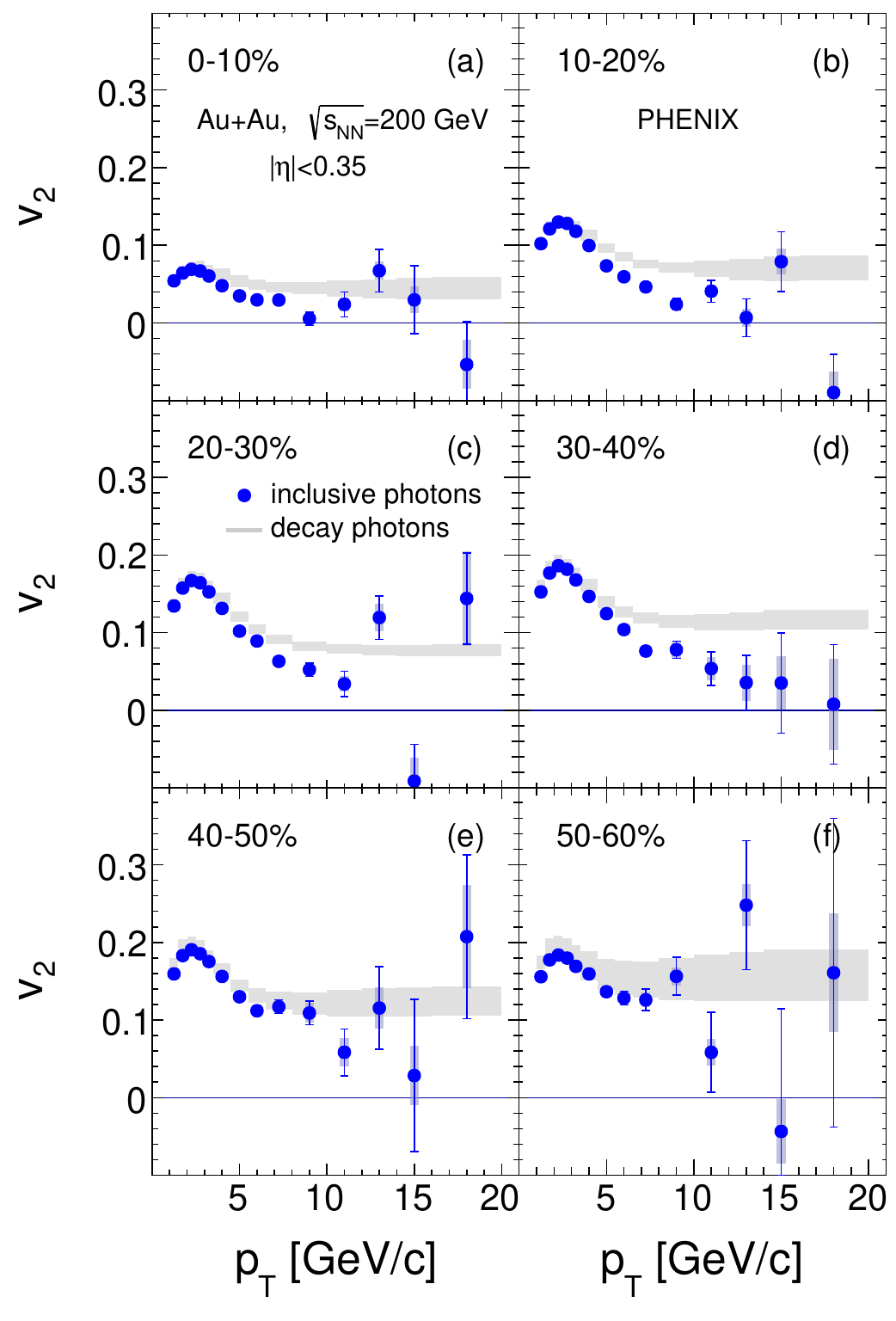}
    \caption{Comparison of the measured anisotropy of inclusive photon 
production \vinc and the modeled anisotropy for photons from hadron 
decay \vdec. Panels (a) through (f) represent 10\% centrality selections 
from central to semiperipheral collisions. Statistical uncertainties on 
\vinc are shown as error bars, while systematic uncertainties are shown 
as shaded bars. For \vdec a band covering the possible systematic 
variations is given. }
    \label{fig:inc_dec_comp}
\end{figure}
\begin{figure}[htb]
    \centering
    \includegraphics[width=1.0\linewidth]{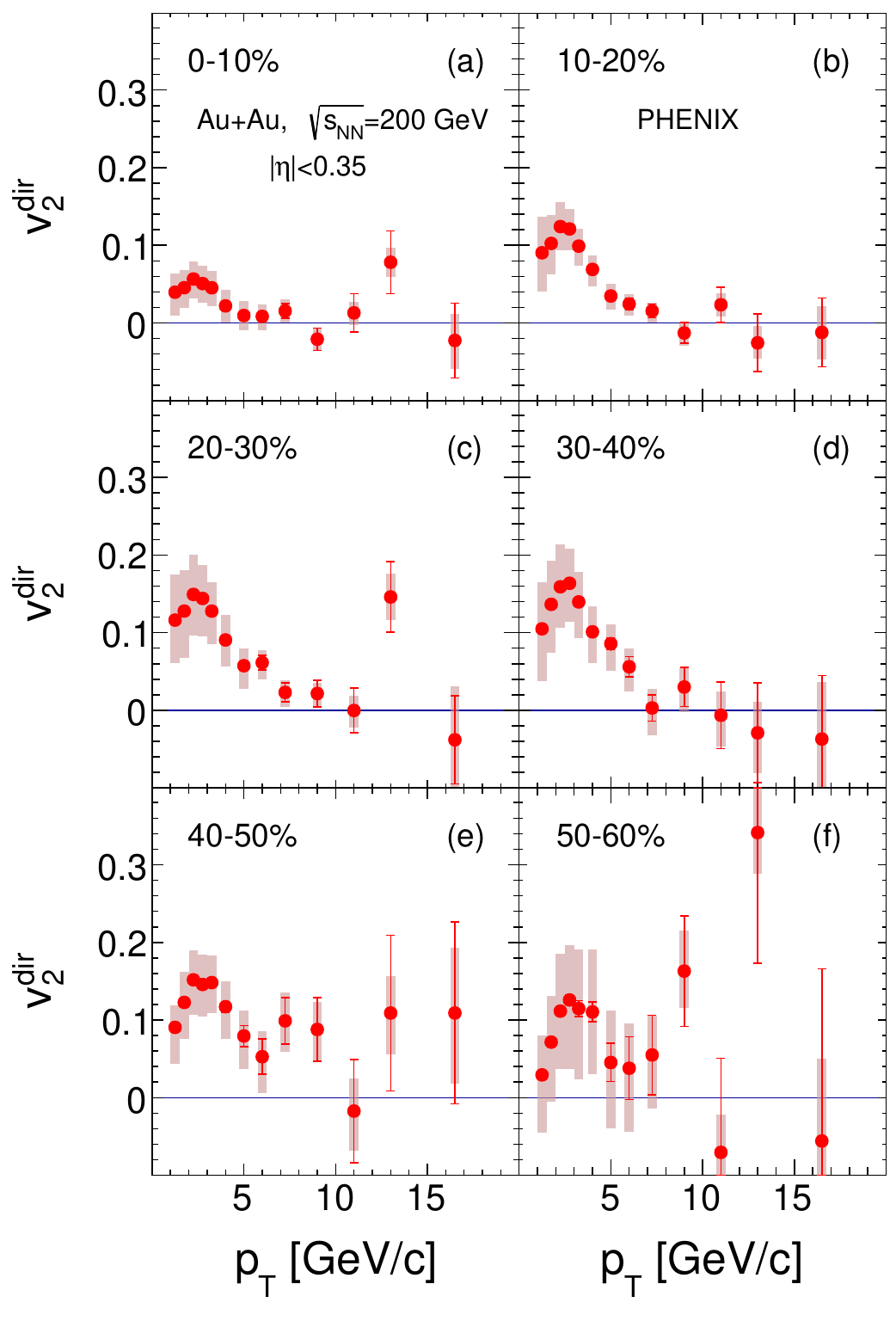}
    \caption{Anisotropy of direct-photon production \vdir. Panels (a) 
through (f) represent 10\% centrality selections from central to 
semiperipheral collisions.  Statistical uncertainties on \vdir are shown 
as error bars, while systematic uncertainties are shown as shaded bars.
}
    \label{fig:dir_results_10}
\end{figure}

\section{\label{sec:sys}Systematic uncertainties}

The systematic uncertainties on \vdir are calculated from the individual 
uncertainties on \vinc, \vdec, and \Rg. Typical values are presented in 
Table~\ref{dir_v2_sys_summary}. There are multiple sources of 
uncertainty on each of these quantities.

The \vinc for the inclusive-photon yield is determined from the yield of 
photon candidates, which is not corrected for detector effects. This is 
based on the premise that the photon reconstruction efficiency does not 
depend on $\Delta\phi$, the angle of the photon with respect to the 
event plane $\Psi_2$, and is independent of the source of the photons. 
The latter is only true if the showers of two decay photons do not 
overlap.

The uniformity and independence of the reconstruction efficiency with 
respect to the reaction plane is estimated by comparing results obtained 
using only the EMCal in the east and west arms respectively. The 
difference in \vinc is 2\%--7\%, depending on \pt and centrality. There 
are also uncertainties on \vinc due to the accuracy with which the event 
plane can be determined. This is estimated by comparing different 
subevent combinations and found to be 1\%--5\%. The effect of shower 
merging is negligible below a photon \pt of 10~\gevc, but the inclusive 
photon yield from 10 to 20~\gevc gradually decreases by up to 10\%, as 
more and more of the showers from $\piz\rightarrow\gamma\gamma$ decay 
photons overlap.

Imperfections due to incomplete suppression of the hadron contamination 
in the inclusive-photon sample are estimated by varying the 
photon-identification cuts. The analysis was repeated using different 
identification cuts and the variation of the results was used as a 
measure to estimate the uncertainty, determined to be $\approx$2\%.

The uncertainties on \vdec are dominated by the uncertainties on the 
determination of $v_2$ for \piz. These are determined through the 
fitting procedure, which uses a Monte-Carlo sampling technique. In this 
approach, each data point is given a Gaussian distribution according to 
its statistical uncertainty, and varied independently. All data points 
of a given data set are also varied simultaneously within their 
correlated systematic uncertainties. The resulting uncertainties,  
shown as bands in Fig.~\ref{fig:inc_dec_comp}, are largest at high 
\pt and for the semiperipheral bin where they reach 25\%. The additional 
uncertainty on $v_2^{\piz}$ arising from the event-plane 
determination is $\approx$3\%. The contributions to \vdec from other mesons 
are small, and the uncertainties introduced by the $KE_T$ scaling are 
estimated to be 2\%. Other uncertainties are negligible.

Finally, the uncertainties on \Rg are estimated through the fitting 
procedure, similar to the process used for the $v_2^{\piz}$ 
uncertainties. An example is given as the band in Fig.~\ref{fig:rgamma}. 
Depending on \pt and centrality the uncertainties are estimated to vary 
from 5\% to 30\%.  Most uncertainties are correlated across different 
centrality ranges. The uncertainty from the event-plane determination is 
a scale uncertainty, while all other uncertainties are correlated in 
\pt.
 
The \vdir is calculated from \vinc, \vdec, and \Rg using 
Eq.~\ref{eqn:vdir}. Due to the nonlinear dependence on \Rg, 
uncertainties are asymmetric unless \vinc = \vdec. To address this, the 
sampling procedure developed in ~\cite{PHENIX:2015igl} was applied. Each 
component of the calculation of \vdir is modeled as a Gaussian 
distribution with its uncertainty as the width. Using these 
distributions, many calculations of \vdir are performed, producing a 
probability distribution for \vdir. Asymmetric systematic uncertainties 
are determined from $\Delta^+_{\vdir}$ and $\Delta^-_{\vdir}$, the 
difference between the two \vdir values where the probability reaches 
half of the maximum value to the \vdir value at the maximum probability.  
The uncertainties are quoted as $2\Delta^\pm_{\vdir}/2.35$.

\section{\label{sec:result}Results}

The values of the inclusive photon anisotropy \vinc and the expected 
anisotropy \vdec of photons from hadron decays are compared in 
Fig.~\ref{fig:inc_dec_comp}, which shows six 10\% centrality classes from 
0\%--10\% to 50\%--60\% of Au$+$Au collisions at \snn{200}. The \pt 
range extends from 1 to 20~\gevc, which significantly increases the 
range previously accessible, \pt$<$ 5~\gevc~\cite{PHENIX:2015igl} and 
10~\gevc~\cite{PHENIX:2011oxq}. For all \pt, \vinc is smaller than 
\vdec, however below 4~\gevc the difference is small and within 
uncertainties. Above \pt$>4$~\gevc, \vinc is visibly lower than the 
\vdec for most cases, signaling a substantial contribution of direct 
photons from a source that does not exhibit an anisotropy with respect 
to the event plane. This is consistent with the expectation that prompt 
photon production would dominate in this \pt range and these photons are 
not expected to have an anisotropy.

The values of \vdir, calculated from \vinc, \vdec, and \Rg using 
Eq.~\ref{eqn:vdir}, are depicted in Fig.~\ref{fig:dir_results_10}. 
For \pt$<4$~\gevc, \vdir is large and very similar to $v_2$ observed for 
pions~\cite{PHENIX:2014uik}. The value increases up to $\approx$2--3~\gevc, 
where it has a broad maximum. While \vdir decreases towards higher \pt, 
nonzero \vdir values seem to extend to well beyond 4~\gevc in \pt. Only 
above 8--10~\gevc, \vdir becomes consistent with zero.

\begin{figure}
    \centering
    \includegraphics[width=1.0\linewidth]{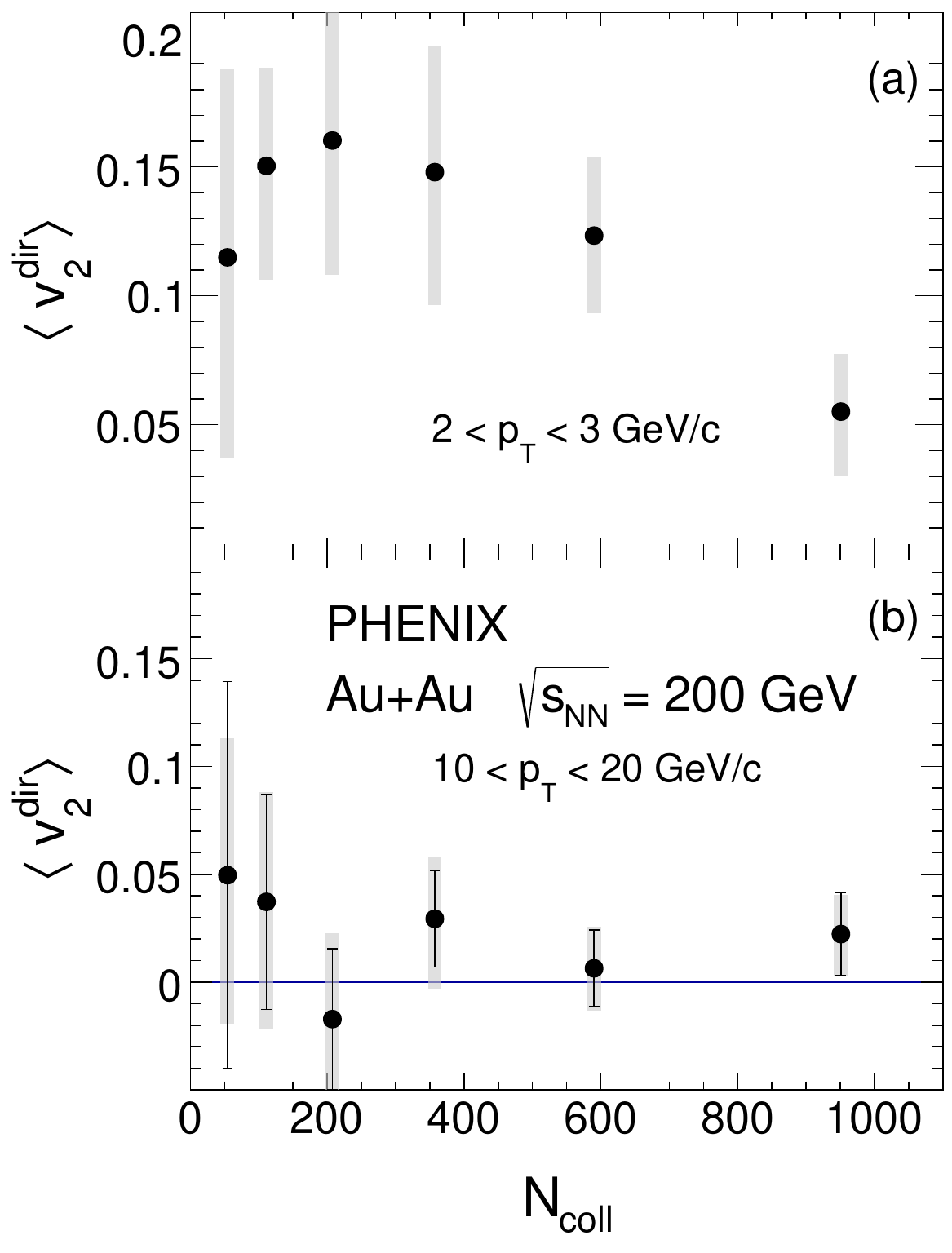}
    \caption{Average azimuthal anisotropy for two different \pt ranges 
as a function of centrality, here represented by the number of binary 
nucleon-nucleon collisions \Ncoll. Panel (a) shows the range from 2 to 3 
\gevc where the anisotropy is maximal. Panel (b) shows the high \pt 
range from 10 to 20~\gevc, where the photon is dominated by prompt 
production through hard scattering processes. For each point statistical 
errors are represented by lines, while systematic uncertainties are 
shown as shaded band.
    }
    \label{fig:centdep}
\end{figure}

The maximum value \vdir reaches is $\approx$0.16 for the semicentral event 
selection of 30\%--40\% centrality or \Ncoll$\approx$200. The centrality 
dependence of the peak value is shown in Fig.~\ref{fig:centdep}(a). 
Although the systematic uncertainties are significant, because they are 
correlated, the data establish a decrease of \vdir towards higher and 
lower values of \Ncoll.  At high \pt$>10$~\gevc, the direct-photon 
anisotropy is consistent with zero over the entire range from central to 
semiperipheral collisions with \Ncoll$\approx50$, as seen in 
Fig.~\ref{fig:centdep}(b).

\begin{figure}
    \centering
    \includegraphics[width=1.0\linewidth]{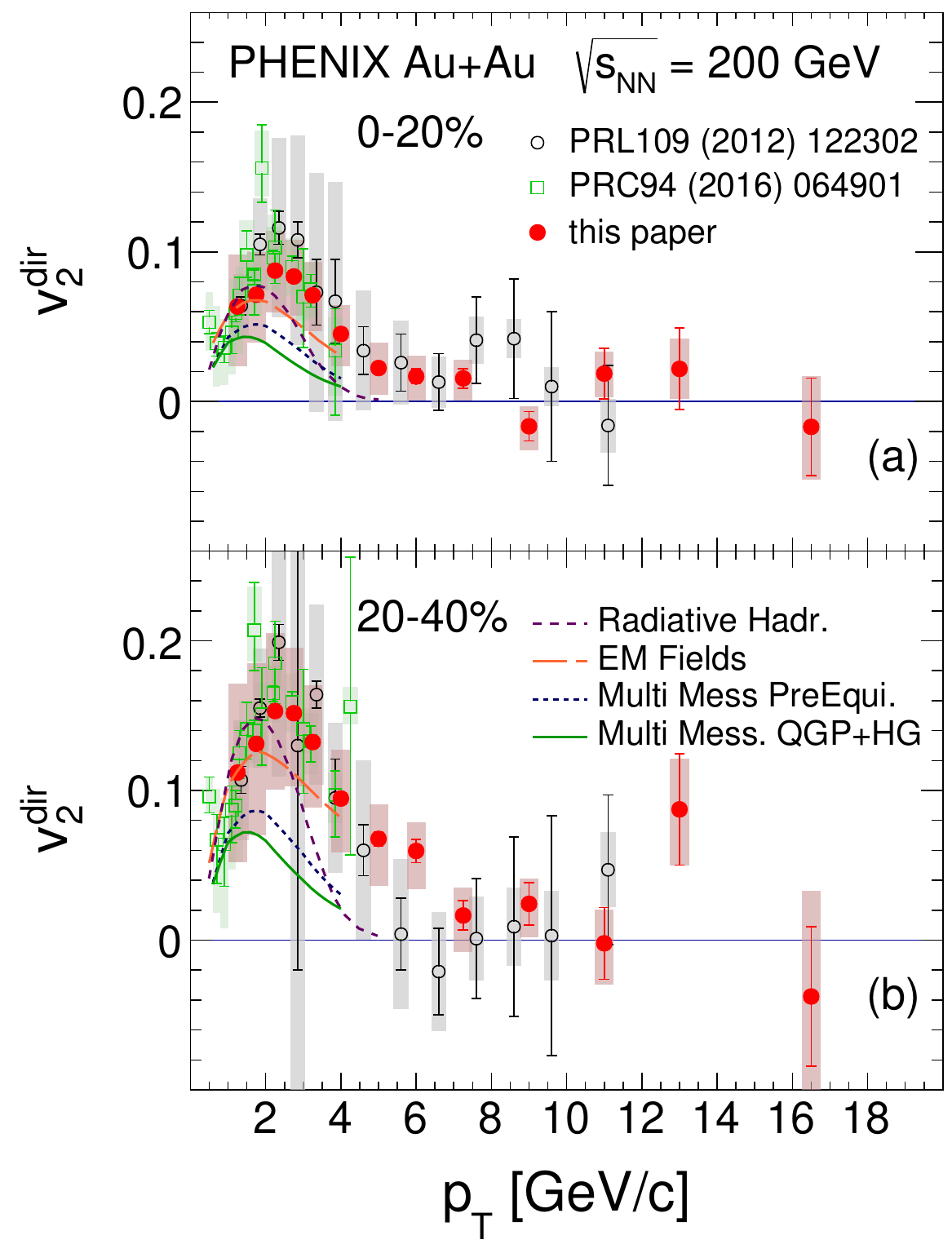}
   \caption{Azimuthal anisotropy for 0\%--20\% (a) and 20\%--40\% (b). 
Again statistical errors are represented by lines, while systematic 
uncertainties are shown as shaded band. Some data points were moved 
slightly in \pt for clarity. The data is compared to previous 
measurements~\cite{PHENIX:2011oxq,PHENIX:2015igl} as well as to 
different theoretical models including pre-equilibrium contributions in 
the multimessenger model~\cite{Gale:2021emg}, modifications to such a 
scenario by the magnetic field present in the reaction 
volume~\cite{Sun:2024vsz}, or radiative contributions at 
hadronization~\cite{Fujii:2022hxa}. }
    \label{fig:model}
\end{figure}

\section{\label{sec:discussion}Discussion and conclusion}

The data are consistent with previous 
publications~\cite{PHENIX:2011oxq,PHENIX:2014uik}, as can be seen in 
Fig.~\ref{fig:model}, where the data are aggregated for 0\%--20\% and 
20\%--40\% centrality and compared directly to previously published 
data. For these two centrality bins also a number of theoretical 
calculations of the direct-photon anisotropy exist, covering the \pt 
range up to 5~\gevc .

The value of \vdir reaches a maximum for all centralizes 
$\approx$2--3~\gevc. Model calculations of thermal photons emitted from 
the rapidly expanding QGP and HG phases~\cite{vanHees:2011vb, 
Linnyk:2015rco, Paquet:2015lta, Shen:2013vja, Chatterjee:2008tp} also 
show a peak structure at similar \pt. A representative calculation of 
the QGP plus HG contributions~\cite{Gale:2021emg}, is shown in 
Fig.~\ref{fig:model} to illustrate this point. However, the values of 
\vdir and the direct-photon yields associated with them are generally 
found to be too low~\cite{PHENIX:2014uik} to explain the data. This 
mismatch is part of the ``direct-photon puzzle''~\cite{David:2019wpt}.

In the multimessenger approach~\cite{Gale:2021emg}, contributions from 
the pre-equilibrium stage have been added to those of the QGP and HG, 
seen in Fig.~\ref{fig:model} as the dotted line. While these 
contributions can increase the anisotropy and seem to extend 
anisotropies further in \pt, the modeled values for \vdir remain too 
low. One possible additional source could be a radiative contribution 
from the hadronization~\cite{Fujii:2022hxa}. Adding this contribution to 
the emission from QGP and HG may achieve a more quantitative description 
of \vdir at low \pt~\cite{Fujii:2022hxa}, given in Fig.~\ref{fig:model} 
as dashed line. However, the \vdir that the radiative hadronization adds 
is limited in its \pt reach to well below 4~\gevc, whereas sizable \vdir 
values are experimentally observed at higher \pt.

Another conjectured source is the modified emission due to the 
electromagnetic fields created in the collision. Various aspects of 
possible modifications have been 
studied~\cite{Basar:2012bp,Muller:2013ila,Wang:2020dsr,Sun:2024vsz}. In 
a recent paper~\cite{Sun:2024vsz} it has been argued that the anisotropy 
of direct-photon emission may increase significantly, even when the 
fields are already substantially reduced in strength. Within the 
systematic uncertainties, considering this effect on top of the 
multimessenger approach with pre-equilibrium contributions leads to a 
promising agreement over the previously measured \pt range up to 
4~\gevc, as seen in the figure. In addition, this source may generate 
substantial \vdir even at higher \pt.  It should be noted, however, that 
PHENIX also measured substantial third order Fourier component $v_3$ for 
direct photons~\cite{PHENIX:2015igl}, which is not discussed 
in~\cite{Sun:2024vsz}, while earlier calculations of magnetic photon 
emission predicted vanishing $v_3$~\cite{Bzdak:2012fr}. The calculation 
of magnetic field effects~\cite{Sun:2024vsz} includes all sources from 
the multimessenger approach~\cite{Gale:2021emg}, but does not consider 
radiative hadronization, while the radiative hadronization 
calculation~\cite{Fujii:2022hxa} ignores pre-equilibrium and magnetic 
effects.  It would be interesting to see the results of a calculation 
with all those conjectured sources included.

\section{\label{sec:summ}Summary}

To summarize, PHENIX presented the measurement of the azimuthal 
anisotropy of direct photons in Au$+$Au collisions at RHIC at 
$\sqrt{s_{_{NN}}}=200$ GeV, extending the range to \pt of 20 \gevc. The 
dependence of the \vdir data on \pt and centrality may help set 
additional constraints and distinguish between proposed sources. For 
example, the contribution from radiative hadronization would be driven 
by the anisotropy and temperature of the source at hadronization. In 
contrast, emission modified by the electromagnetic fields would be 
sensitive to earlier times, higher temperatures, and the centrality 
dependence of the strength of the created fields.

Finally, at high \pt$>10$~\gevc, the direct-photon anisotropy is broadly 
consistent with zero for all centralizes. Unlike \piz or jets, prompt 
photons are not quenched, and thus should be unaffected by the collision 
geometry. The absence of an anisotropy supports the interpretation that 
direct photons in this \pt region are predominantly prompt photons from 
initial hard scatterings.


\begin{acknowledgments}

We thank the staff of the Collider-Accelerator and Physics
Departments at Brookhaven National Laboratory and the staff of
the other PHENIX participating institutions for their vital
contributions.  
We acknowledge support from the Office of Nuclear Physics in the
Office of Science of the Department of Energy,
the National Science Foundation,
Abilene Christian University Research Council,
Research Foundation of SUNY, and
Dean of the College of Arts and Sciences, Vanderbilt University
(U.S.A),
Ministry of Education, Culture, Sports, Science, and Technology
and the Japan Society for the Promotion of Science (Japan),
Natural Science Foundation of China (People's Republic of China),
Croatian Science Foundation and
Ministry of Science and Education (Croatia),
Ministry of Education, Youth and Sports (Czech Republic),
Centre National de la Recherche Scientifique, Commissariat
{\`a} l'{\'E}nergie Atomique, and Institut National de Physique
Nucl{\'e}aire et de Physique des Particules (France),
J. Bolyai Research Scholarship, EFOP, HUN-REN ATOMKI, NKFIH,
and OTKA (Hungary), 
Department of Atomic Energy and Department of Science and Technology (India),
Israel Science Foundation (Israel),
Basic Science Research and SRC(CENuM) Programs through NRF
funded by the Ministry of Education and the Ministry of
Science and ICT (Korea).
Ministry of Education and Science, Russian Academy of Sciences,
Federal Agency of Atomic Energy (Russia),
VR and Wallenberg Foundation (Sweden),
University of Zambia, the Government of the Republic of Zambia (Zambia),
the U.S. Civilian Research and Development Foundation for the
Independent States of the Former Soviet Union,
the Hungarian American Enterprise Scholarship Fund,
the US-Hungarian Fulbright Foundation,
and the US-Israel Binational Science Foundation.

\end{acknowledgments}


%
 
\end{document}